\documentclass[prd,preprintnumbers,twocolumn,aps]{revtex4-1}
\usepackage[english]{babel}
\usepackage{graphicx}
\usepackage{dcolumn}
\usepackage{bm}
\usepackage{epsfig,amsmath,mathrsfs}
\usepackage{amssymb}
\usepackage{hyperref}
\usepackage{xcolor}
\hypersetup{final}

\def \nn{\nonumber}

\begin{document}

\title{Acoustic Hawking radiation from an evolving horizon in a dynamical analogue spacetime}

\author{Oindrila Ganguly}
\email{oindrilag@iisc.ac.in, lagubhai@gmail.com}
\affiliation{Indian Institute of Science, Bengaluru 560012, India}

\begin{abstract}
Our knowledge of dynamical black holes suffers from a lack of observational insight. In an analogue model of gravity, we can design a longitudinally symmetric dynamical acoustic black hole with a moving horizon. In this symmetric spacetime, the marginally outer trapped surface and the so called evolving horizon are degenerate. Interestingly, there are many ways of assigning a surface gravity to the horizon in the absence of time translation invariance. Here, we present two of them that are distinguished by whether they are defined only locally or take into account  the global properties of spacetime outside the black hole. It is expected that a dynamical black hole would emit spontaneous thermal radiation and its temperature would be proportional to a surface gravity of the moving horizon but there is no consensus on which of the surface gravities would give the Hawking temperature. We propose that a non-stationary analogue spacetime when realised experimentally can possibly help resolve these mysteries and provide the first observational signatures of dynamical Hawking radiation. 
\end{abstract}

\maketitle

\section{Introduction}

Analogue models of gravity enable people to experimentally study geometric aspects of strong gravity, the most notable among them being radiation from black holes. Demonstration of correlation between Hawking particles and their partners beyond the acoustic horizon \cite{steinhauer2015,steinhauer2016} in a Bose Einstein condensate, match of the measured Hawking temperature with the theoretically predicted value  \cite{denova2018} and observation of superradiance of acoustic perturbation in gravity waves in water from a rotating acoustic black hole \cite{torres2016} are remarkable successes of this programme. Almost all the theoretical and laboratory based studies of analogue black holes though focus solely on stationary geometries. Real physical black holes are rarely stationary as they accrete matter and energy, undergo evaporation by spontaneous or stimulated emission and merge with neutron stars or other black holes. Such dynamical black holes can be studied with ease in analogue systems by making either the velocity of the medium or the speed of relevant perturbations or both time dependent. It is expected that spontaneous emission of particles by Hawking radiation will  continue near the horizon of a dynamical black hole \cite{visser2001,hayward2008,kinoshita2011}. 

From a theoretical point of view, in such highly dynamical situations, the standard description of black holes in terms of an event horizon in a stationary, asymptotically flat spacetime is ill suited. It becomes impossible to identify the event horizon owing to its very definition: in a spacetime manifold $M$, it is defined as the boundary of the causal past of future null infinity, that is, $\mathscr H = \partial(J^-(\mathscr I^+))$ and the black hole is defined as  $B = M - J^-(\mathscr I^+)$. Thus, the event horizon is a global causal concept and ideally one has to wait infinitely long, except in stationary spacetimes, to ascertain its location, if it exists at all. Thus, to describe a dynamical black hole, it is important to have (quasi) locally defined, geometric ideas of horizons. Various locally defined horizons have been proposed in the literature, like Hawking's apparent horizon \cite{hawkingellis}, Ashtekar and his collaborators' isolated and dynamical horizons \cite{ashtekar2004}, Hayward's trapping horizons and marginally trapped surfaces \cite{dicriscienzo2009,hayward1997,hayward1993a}, and Nielsen and Visser's evolving horizon \cite{nielsen2005}. An issue with local horizons is that they are not unique, unlike the event horizon, and their existence depends on the nature of foliation/slicing of spacetime. There may exist particular asymmetric foliations of spacetime that do not admit any horizon while a symmetric slicing may have one or all of them and they may even coincide.  On the other hand, a spacetime may have many marginally trapped surfaces and trapping horizons. 
There is no theoretical consensus on which local horizon may be taken to define a black hole, which one is the horizon where laws of black hole thermodynamics hold and from which of these does Hawking radiation occur. It is not even known whether one horizon meets all these requirements. Detailed discussion on them is available in  \cite{nielsen2008} and references therein. Understandably, observational feedback would be of immense help in this conundrum and this can only be provided by analogue models of gravity at the moment. 
Hawking radiation is a purely local, kinematical effect associated with the behaviour of vacuum fluctuations of a quantum field in a general Lorentzian geometry with a `horizon' \cite{visser1997,barcelo2006} and is expected to be robust under time dependence of the geometry.

In this article, we study general features of a time dependent acoustic geometry that can be manifest in any analogue model. The actual system may be a Bose Einstein condensate or an evolving quark-gluon plasma or any other quantum fluid, seen in the hydrodynamic limit. Usually, spherical symmetry is assumed while studying dynamical spacetimes without angular momentum in gravity. However, non-rotating acoustic spacetimes  typically inherit a slab or cylindrical geometry from the nature of flow of the underlying fluid. This gives rise to a few qualitative distinctions between the two. Two simplifying assumptions that we here make are that the fluid has a non-trivial velocity only in one direction which we call the longitudinal axis and the velocity, pressure and density of the fluid are  functions only of time and position along this longitudinal axis. Thus, we have a longitudinally symmetric system. This actually is a realistic assumption for analogue systems as found in the experimental set up of  \cite{steinhauer2015,steinhauer2016,denova2018}. The emergent acoustic spacetime inherits this symmetry and we further choose to slice the spacetime at any instant of time in such a way that the spacetime foliations are planar. Apart from this, the fluid needs to be barotropic and its velocity irrotational to give rise to an effective acoustic spacetime. In section \ref{sec:metric}, we show that the acoustic metric then is naturally like the metric of a spherically symmetric  dynamical spacetime in Painlev\'{e}-Gullstrand (PG) coordinates \cite{visser1998}, but for the flat geometry of the constant time spatial hypersurfaces. We find that, in this system the evolving horizon is degenerate with the  marginally (outer) trapped surface.  In section \ref{sec:surfgr}, we evaluate  dynamical surface gravities of the evolving horizon using a local and a non-local definition of the same. These two definitions do not agree on the value of surface gravity at the moving horizon. Their differences can be intuitively understood by thinking in terms of physical gravitational black holes.  The non-local version of the definition actually takes into account how the distribution of matter outside the black hole changes with time as opposed to the local one. Only in the special case of a stationary black hole in vacuum, do they become same. We know that for such a black hole, Hawking temperature is proportional to the surface gravity of the horizon. Naturally, the dynamical situation is much more nuanced - firstly because there is no unique local horizon and secondly because even in spacetimes with high symmetry that make the local horizons merge, there are many ways to compute the surface gravity. So, the question remains which surface gravity is directly related to Hawking temperature. We shall discuss this in section  \ref{sec:surfgr}.

\section{Acoustic dynamical metric, evolving horizon and trapped surfaces} \label{sec:metric}

To begin with, we choose the spatial coordinates $\mathbf{x} = (x,y,z)$ to describe the system in such a way that the bulk three velocity of the fluid   $\mathbf{v}_0^\alpha(t,\mathbf{x})=(v_0(t,x),0,0)^\alpha$ 
\footnote{We use lowercase Latin alphabets as spacetime indices and lowercase Greek alphabets as spatial indices only.}.
Here, $t$ is the time measured by the laboratory observer. Thus, the $x$ axis is aligned along the direction of flow so that $\mathbf{v}_0(t,\mathbf{x}) = \mathbf{v}_0(t,x)$ and the fluid flows towards decreasing $x$. The $y$ and $z$ axes span the transverse plane. The bulk variables describing the fluid are its velocity $\mathbf v_0(t,x)$, pressure $p_0(t,x)$ and density $\rho_0(t,x)$. We shall, henceforth, avoid writing the suffix $`0'$ as there is no risk of confusion. Remember, we have assumed longitudinal symmetry which means that all the physical variables are functions only of $(t,x)$ and not of $y$ and $z$. Now, we need to make sure that this general fluid meets the prerequisite criteria necessary for the construction of an analogue model. Since, we have in the back of our mind a Bose Einstein condensate or a general superfluid as the model system, we can safely take the fluid to be inviscid. Additionally, the fluid must follow a barotropic equation of state, that is, $p=p(\rho)$. Our chosen form of the velocity field $\mathbf{v}(t,x)$ is, by construction, locally irrotational. Then, we can straight away write down the acoustic metric $g$ in terms of the acoustic spacetime interval as \cite{unruh1981,visser1998}:
\begin{align}
ds^2
&= 
\frac{\rho(t,x)}{c_s(t,x)}
\Big[
-(c_s(t,x)^2-v(t,x)^2) dt^2
+ 2 v(t,x) dt dx              \nn \\
& \ + dx^2 + dy^2 + dz^2
\Big]~.
\label{eq:acmet}
\end{align}
Here, $c_s(t,x)$ is the local speed of sound defined by $c_s^2=\partial p/\partial \rho$. This form of the acoustic metric is similar to the metric of a general dynamical, spherically symmetric physical spacetime expressed in Painlev\'{e}-Gullstrand coordinates $(\tilde t, r, \theta, \phi)$ \cite{nielsen2005}
\begin{align}
ds_{PG}^2 
&=
- \left(
c(\tilde t, r)^2 - v(\tilde t, r)^2
\right) dt^2
+ 2 v(\tilde t, r) dt dr 
+ dr^2				\nn \\
& \ + r^2 d\Omega^2~.
\label{eq:pgmet}
\end{align}
The only difference is that our acoustic metric of \eqref{eq:acmet} has longitudinal symmetry instead of spherical symmetry assumed in deriving eq. \eqref{eq:pgmet} and a spacetime dependent conformal factor. The future directed `outgoing' and `ingoing' longitudinal null curves in this acoustic spacetime have $ds^2=0$ together with $dy=dz=0$. Thus, along them,
\begin{align}
\frac{dx}{dt}=-v \pm c_s~.
\label{eq:null}
\end{align}
The `$+$' and `$-$' signs refer to outgoing and ingoing null curves respectively.
The `outward' and `inward' pointing null vectors $l$ and $n$ tangent to the corresponding null curves are:
\begin{align}
l^a =
\alpha (t,x)
(1,-v(t,x)+c_s(t,x),0,0)^a 
\label{eq:l}
\end{align}
and
\begin{align}
n^a =
\beta(t,x)
(1,-v(t,x)-c_s(t,x),0,0)^a~,
\label{eq:n}
\end{align}
where $g(l,l)=g(n,n)=0$ and $\alpha(t,x), \beta(t,x)$ are scaling functions. The cross-normalisation between $l$ and $n$ is $g(l,n) = -2 \alpha\beta c_s^2$. The (future directed) inward pointing null curves are always seen to be ingoing from eq. \eqref{eq:null} and eq. \eqref{eq:n}, irrespective of the relative values of $v$ and $c_s$. However, when $v(t,x) > c_s(t,x)$, we have $dx/dt < 0$ which makes $l^x<0$  i.e. the `outward' pointing null curve also becomes ingoing. This tells us that we can define an acoustic horizon by the condition
\begin{align}
c_s(t,x)=v(t,x)~,
\label{eq:achor}
\end{align}
since, in the region having $v(t,x)>c_s(t,x)$, all acoustic perturbations are carried by the flowing fluid away from the evolving horizon, towards lower values of $x$. So, no acoustic perturbation can escape the region having $v(t,x)>c_s(t,x)$, bounded by the acoustic horizon. This implicitly defines a function $x_h(t)$ giving the time dependent location of the horizon:
\begin{align*}
c_s(t,x_h(t))=v(t,x_h(t))~.
\end{align*}
The position of the acoustic horizon is no longer constant in time, thus giving a moving or evolving horizon. This notion of a moving horizon that naturally comes about in time dependent analogue models of gravity is seen used  to formally define an ``evolving horizon" for a general dynamical spacetime in \cite{nielsen2005}. Surfaces on which $c_s (t,x) < v(t,x)$ form outer trapped surfaces and they together constitute the outer trapped region. The outer boundary of the outer trapped region is the marginally outer trapped surface (MOTS) or apparent horizon and is defined by the same condition as that of eq. \eqref{eq:achor}. So, in our model, the moving/evolving horizon and apparent horizon coincide.

\section{Dynamical surface gravity and Hawking radiation} \label{sec:surfgr}

The surface gravity is a geometrical quantity determined by  local geometry at the horizon. It may also depend on non-local information provided through asymptotic normalisation conditions. In a static asymptotically flat spacetime, the event horizon is also a Killing horizon and the surface gravity is defined in terms of the corresponding Killing vector field. This Killing definition of surface gravity is non-local since the normalisation of the Killing vector field is fixed in such a way that it concurs with the four velocity of an asymptotic observer at rest. When the spacetime becomes dynamical, there is no longer any timelike Killing vector field. (This also implies that, thinking in terms of a physical black hole, we are looking at a non-vacuum solution.) In the dynamical case, there are several (at least five are known) inequivalent definitions of surface gravity - they can be distinguished by whether they depend on some specific symmetry of the system (like Hayward's definition in terms of Kodama vector that is best suited to spherically symmetric spacetimes), whether they are local or non-local and the varied choices of normalisation made. To understand how the various versions/definitions of surface gravity compare with each other, please refer to  \cite{pielahn2011}. They all merge in the special case of a static spherically symmetric vacuum black hole but do not all coincide when the black hole exterior is not vacuum. 

For the simple dynamical acoustic spacetime being considered here, we present definitions of surface gravity in terms of the future directed outgoing and ingoing null normals $l$ and $n$. Remember, we had left the scaling functions $\alpha(t,x)$ and $\beta(t,x)$ unspecified while writing the component functions of $l$ and $n$. Different choices of these scaling functions correspond to different normalisations and consequently distinct definitions of surface gravity. Below we shall explore two definitions - the first one is due to Nielsen and Visser \cite{nielsen2005} and the second one is the adaptation of Fodor et al's definition to PG coordinates \cite{pielahn2011} applied to our acoustic spacetime. Both these definitions rest on the fact that the null geodesics, in particular the outgoing null geodesics are non-affinely parameterised. The surface gravity is a measure of this inaffinity:
\begin{align}
(\nabla_l l)^a 
=
\kappa l^a~,  \label{eq:kappa} 
\end{align}
The difference between $\kappa_{NV}$ and $\kappa_F$, proposed by Nielsen-Visser and Fodor et al respectively lie in the choices of the scaling functions. Nielsen and Visser have chosen the cross-normalisation $g(l_{NV},n_{NV})=-2$ and the most symmetric choice of normalisation where $\alpha (t,x) = \beta (t,x) = 1/c_s (t,x)$. (In the equations appearing below, we will often suppress the explicit functional dependence of $c_s$ and $v$ on $(t,x)$, for brevity. We shall also use a dot and a prime to denote differentiation with respect to $t$ and $x$ respectively.) Adopting this, we get 
\begin{align*}
\kappa_{NV} (t,x)
=
\frac{c_s' (t,x)-v' (t,x)}{c_s(t,x)}~.
\end{align*}
On the moving horizon, 
\begin{align}
\kappa_{NV,h} (t,x_h)
= 
\left[
\frac{1}{c_s} 
(c_s^\prime - v^\prime)
\right]_{x=x_h}~.
\label{eq:kappanv}
\end{align}
This is a local definition of surface gravity as the scaling functions of $l_{NV}$ and $n_{NV}$ are chosen completely locally. Though Fodor et al's choice of the scaling function is best suited to the ingoing Eddington-Finkelstein (EF) coordinates, we here use the adaptation of their definition to PG coordinates presented in \cite{pielahn2011}. Here, $g(l_F,n_F)=-1$ and $\alpha(t,x)=1$. This choice is special because at the evolving horizon, $l_F \rightarrow t$ where $t^a=(1,0,0,0)$. The vector field $t$ would have been a timelike Killing vector field if the spacetime were static. Even if the spacetime were only asymptotically static and flat, $t$ would be the four velocity of the asymptotic observer at rest. Since, $l_F=c_s l_{NV}$, it is straightforward to show using eq. \eqref{eq:kappa} that,
\begin{align*}
\kappa_F (t,x) & \, = c_s(t,x)\kappa_{NV}(t,x) + l_{NV}^a \nabla_a c_s (t,x) \\
& \, = (c_s^\prime - v^\prime) + \frac{\dot{c_s}}{c_s} + (c_s - v)\frac{c_s^\prime}{c_s}~.
\end{align*}
On the evolving horizon,
\begin{align}
\kappa_{F,h}(t,x_h) = \left(c_s^\prime - v^\prime + \frac{\dot{c_s}}{c_s} \right) \Bigg |_{x=x_h}~.
\label{eq:kappaf}
\end{align}
The longitudinally symmetric constant time slicing of the spacetime ensures that the evolving horizon lies at a fixed $x_h$. Hence, at any  instant $t$, both $\kappa_{NV,h}$ and $\kappa_{F,h}$ remain constant over the horizon. 

The physical significance of surface gravity is that it gives the temperature of Hawking radiation emanating from the horizon of a stationary black hole upto some dimensional constants. We know that Hawking radiation is intimately linked to the presence of a horizon and the property of a horizon that is of relevance here is that it is a one way membrane. Local horizons do share this feature. So, we should expect to have dynamical Hawking radiation. A point of concern is whether this radiation is thermal. It seems logical to expect that as long as the time scale over which the energy of the black hole changes is larger than the time taken by light to travel half way across the horizon, the horizon would be able to settle into its new state and would have a temperature. The Hawking flux emanating from the horizon would also then be thermal and this temperature should be related to an appropriate dynamical surface gravity of the horizon. Thus, we see that the primary hurdle to our understanding of spontaneous radiation from dynamical black holes lies in the non-uniqueness of local horizons and the dynamical surface gravity. In \cite{hayward2008}, Hayward et al have applied a Hamilton Jacobi variant of the Parikh Wilczek tunnelling method to determine a local redshift-renormalised Hawking temperature $T = \kappa_{Kod,h}/2\pi$, where $\kappa_{Kod}$ is the surface gravity defined using  Kodama vector for any future outer trapping horizon of a non-stationary spherically symmetric black hole. On the other hand, in \cite{visser2001}, Hawking radiation from a slowly evolving apparent horizon in a spherically symmetric Lorentzian spacetime has been shown to occur with a temperature $T = \kappa_{V,h}/2\pi$ where $\kappa_V$ is a surface gravity proposed by Visser. In \cite{kinoshita2011}, the observable Hawking temperature of an eternal black hole with slowly changing mass is found to be associated with the inaffinity of ingoing null geodesics at or near the past horizon. In reference \cite{barcelo2010a}, the authors have gone a step further to construct a general framework for getting Hawking-like fluxes for a broad class of spacetimes including those where no horizon forms, like those of a collapsing object. The only resolution between these differing propositions can be achieved through observational input coming from analogue systems.

\section{Observational prospects and concluding remarks} \label{sec:concl}

The surface gravity of a horizon does not have any direct observational manifestation. But acoustic Hawking radiation can be observed and Hawking temperature measured, as has been done by Steinhauer and his collaborators for a static acoustic apparent horizon in an atomic Bose Einstein condensate (BEC) \cite{steinhauer2015,steinhauer2016,denova2018}. Detailed studies of spectra and correlations of `Hawking phonons' in transonic \textit{stationary} flows of dilute atomic condensates have been presented in articles like \cite{lahav2009,macher2009,recati2009,michel2016} and references therein. In the new experiment reported in ref. \cite{kolobov2021}, the authors have observed that after the formation of an acoustic horizon in a BEC, there is an initial ramp-up of Hawking radiation after which it attains the phase of spontaneous, stationary thermal emission, like real stationary black holes formed by gravitational collapse.

We are interested in a purely dynamical analogue black hole as it would mimic a black hole that is growing or shrinking. This may be achieved by considering a flowing condensate characterised by a time dependent interaction constant in a time dependent potential. This would make both the speed of sound in the condensate and the phase velocity vary with time. An estimation of the temperature of spontaneously emitted phonons in this case can possibly help us identify the `correct' definition of dynamical surface gravity.  Another very interesting phenomenon that comes about in a time dependent spacetime is the dynamical Casimir effect (DCE). In this process, real particles are produced from vacuum excitations of a quantum field in a time dependent geometry. A question that arises in this case is how does one distinguish between particle creation due to dynamical Casimir effect and dynamical Hawking radiation in an analogue experiment. However, remember that Hawking radiation occurs only in the vicinity of the horizon while dynamical Casimir effect gives rise to creation of particles over the entire extent of the system. This has also been observed in numerical simulations of Hawking radiation from acoustic black holes \cite{carusotto2008}, where dynamical Casimir effect occurs transiently when the velocity of the fluid is changed.

Furthermore, with the freedom that  analogue models allow, we may successfully design acoustic spacetimes where different local horizons are non-degenerate and even non-unique. This could permit us to investigate which of the local geometric horizons actually defines a black hole and whether the same also participates in Hawking radiation. True, that in analogue systems, we can only study the kinematical aspects of general relativity and there is no analogue of black hole thermodynamics, still it holds enough promise to play a pivotal role in revealing at least a part of the physics of  black holes in time dependent spacetimes.

\section*{Acknowledgement}

I thank Chaitra Hegde for numerous discussions about particle production in a dynamical spacetime . I also thank Ajit Srivastava and Subhendra Mohanty for their comments on the manuscript, Bhavesh Chauhan  for asking an important question and Navinder Singh for helping me with confusions about Bose Einstein condensates.

\bibliographystyle{apsrev4-1}
\bibliography{an_gr,an_bec,dynhor}

\end{document}